\documentclass[sn-mathphys-num,iicol]{sn-jnl}
\usepackage{graphicx}

\usepackage{amsmath}
\usepackage{braket}
\usepackage{amsfonts}

\usepackage{booktabs}

\usepackage{aas_macros}

\usepackage{caption}
\usepackage{graphicx}
\usepackage{subcaption}

\usepackage{bm}
\usepackage{slashed}

\usepackage{changes}

\def\aap{\textit{Astronomy and Astrophysics}}

\def\be{\begin{equation}}
\def\ee{\end{equation}}
\def\bea{\begin{eqnarray}}
\def\eea{\end{eqnarray}}

\def\L{\mathcal{L}}
\def\Lm{\mathcal{L}_m}

\newcommand\rs{\mathit{r}_S}

\definechangesauthor[name={Olivier}, color=magenta]{OM}
\definechangesauthor[name={Aurelien}, color=cyan]{AH}

\begin{document}
\title[Schwarzschild black-hole immersed in an electric or magnetic background in Entangled Relativity]{Schwarzschild black-hole immersed in an electric or magnetic background in Entangled Relativity}

\author*[1,2]{Olivier Minazzoli}\email{ominazzoli@gmail.com}
\author[3,1]{Maxime Wavasseur}
\affil[1]{Artemis, Universit\'e C\^ote d'Azur, CNRS, Observatoire C\^ote d'Azur, BP4229, 06304, Nice Cedex 4, France}
\affil[2]{Bureau des Affaires Spatiales, 2 rue du Gabian, 98000  Monaco}
\affil[3]{Departament de Física Quántica i Astrofísica (FQA), Universitat de Barcelona (UB), Carrer de Martí i Franquès, 1, Barcelona, 08028, Spain}
\abstract{In this paper, we present the solution for a Schwarzschild black-hole immersed in an electric or magnetic background field à la Melvin within the framework of Entangled Relativity. Previous solutions in Entangled Relativity required black-holes to be charged for the matter field to be defined everywhere. This is because the theory precludes the existence of vacuum solutions, thereby satisfying Einstein's definition of Mach's Principle. The current black-hole solutions represent the first exact and neutral black-hole solutions of Entangled Relativity discovered to date. The Schwarzschild black-hole of General Relativity emerges as a limit of these solutions when the background field approaches zero, whereas the Melvin solution of General Relativity does not emerge as a limit when the black hole's size approaches zero. This finding suggests that astrophysical black-holes in Entangled Relativity are indistinguishable from those in General Relativity, given the generally weak interstellar density of matter fields.}
\maketitle

\section{Introduction}

Entangled Relativity is a reformulation of Einstein's General Theory of Relativity that posits a very specific non-linear coupling between the Ricci scalar curvature and matter fields, instead of the usual linear coupling found in Einstein's theory \cite{ludwig:2015pl}. This non-linear coupling prevents the theory from being defined without matter fields from the onset, ensuring that Entangled Relativity satisfies Einstein's original requirement that the metric field is wholly determined by matter, such that no vacuum solution should exist in a satisfying theory of relativity \cite{einstein:1918an,einstein:1918sp}. This requirement for a relativistic theory is known as (Einstein's definition of) Mach's principle \cite{einstein:1918an,einstein:1918sp,pais:1982bk,hoefer:1995cf,minazzoli:2024pn}. Because matter and spacetime curvature are intertwined at the level of the theory's formulation, it has been named `Entangled Relativity' \cite{arruga:2021pr}.

Like General Relativity, Entangled Relativity is an $f(R,\Lm)$ theory \cite{harko:2010ep,Harko_Lobo_2018,harko:2014ga}. However, it stands out as unique within its class because, to the best of our knowledge, it is the only theory that requires fewer parameters than General Relativity to be defined, while still recovering the phenomenology of General Relativity in a generic limit \cite{minazzoli:2018pr,arruga:2021pr,minazzoli:2021cq,minazzoli:2022ar,minazzoli:2023ar}---see Sec. \ref{sec:gen_limit}. As such, it is arguably more consistent with Occam's principle than General Relativity—unlike most, if not all, other $f(R,\Lm)$ theories. 


Indeed, a major motivation for studying Entangled Relativity is that it requires fewer fundamental universal dimensionful constants than standard physics, while still recovering standard physics in a generic limit.\footnote{While standard physics relies on three fundamental universal dimensionful constants, $G$, $\hbar$, and $c$, Entangled Relativity, in a relevant quantum field theory framework depicting the entire scope of physics, requires only two: $\epsilon$ and $c$, where $\epsilon$ is the reduced Planck energy \cite{minazzoli:2022ar,minazzoli:2023ar}---see Eq. (\ref{eq:ERPI}).} Thanks to a cancellation in the field equations—originally found in scalar-tensor theories and named ‘\textit{intrinsic decoupling}’ in \cite{minazzoli:2013pr}—the phenomenology of the theory is surprisingly similar to that of General Relativity in many relevant situations, such as in the solar system \cite{minazzoli:2013pr} and with neutron stars \cite{arruga:2021pr,arruga:2021ep}. The issue of the accelerated expansion of the universe remains unclear within this framework, although there are already a few proposals \cite{minazzoli:2018pr,minazzoli:2021cq,minazzoli:2024ar}. Entangled Relativity naturally converges toward General Relativity without a cosmological constant during the universe's expansion. However, the structure of the theory prevents adding a cosmological constant by hand if one wants Newton's constant, $G$, to remain constant in the dark energy cosmological era \cite{minazzoli:2018pr,minazzoli:2021cq,minazzoli:2024ar}. Therefore, in Entangled Relativity, the acceleration of the expansion of the universe cannot simply be accounted for by a cosmological constant but must originate from something else. Hence, if the theory provides a better depiction of the laws of physics than General Relativity, the $\Lambda$-CDM cosmological model cannot be the most accurate cosmological model possible.

Since the theory prohibits vacuum solutions a priori, the vacuum solutions of General Relativity are not solutions of Entangled Relativity. However, we now have increasing evidence, notably through the observation of the shadows of supermassive black-holes \cite{EHT:2019aj,EHT:2019al} and the ringdown of gravitational waves from the merger of two black-holes \cite{isi:2019pr,LVK:2019pd,LVK:2021pr}, that General Relativity's black-holes provide a good description of very massive compact objects in our universe. Therefore, it is necessary to verify that black-hole solutions in Entangled Relativity do not deviate significantly from those in General Relativity under astrophysical conditions, and/or to predict potential observational deviations that might be expected from Entangled Relativity.

%
To investigate this issue, previous studies have focused on charged black-holes \cite{minazzoli:2021ej,wavasseur:2024ar}. This approach ensures that the matter field—here, the electromagnetic field—is defined everywhere in the solution. These studies notably showed that the Schwarzschild black-hole solution is a limit of the charged black-hole of Entangled Relativity in the vanishing charge limit, which corresponds to the vanishing matter field limit in this case. Since astrophysical black-holes are not expected to have significant charges—either because they are not formed with substantial charges or because they tend to discharge over time by attracting more particles with opposite charges—this result suggests that Entangled Relativity black-holes may be indistinguishable from those of General Relativity under astrophysical conditions.

In the work presented in this communication, we used an alternative strategy. Instead of looking for charged black-hole solutions, we sought neutral (spherical) solutions immersed in either an electric or a magnetic field. The results of these investigations are presented below.

\section{Field equations}

The quantum phase that defines the theory is given by \cite{minazzoli:2022ar,minazzoli:2023ar}
\be
\Theta = -\frac{1}{2 \epsilon^2} \int d^4_g x \frac{\L^2_m(f,g)}{R(g)}, \label{eq:ERPI}
\ee
where $R$ is the usual Ricci scalar that is constructed upon the metric tensor $g$, $ \mathrm{d}^4_g x := \sqrt{-|g|}  \mathrm{d}^4 x$ is the spacetime volume element, with $|g|$ the metric $g$'s determinant, and $\L_m$ is the Lagrangian density of matter fields $f$---such as gauge bosons, fermions and the Higgs---which could be the current \textit{standard model of particle physics} Lagrangian density, but most likely a completion of it. It also depends on the metric tensor, a priori through to the usual \textit{comma-goes-to-semicolon rule} \cite{MTW} in order to recover General Relativity in some limit.\footnote{Strictly speaking, this condition is only necessary in some limit of the theory, but could perhaps be relaxed in general, as long as it then emerges in the required limit.} Let us note that, like General Relativity, Entangled Relativity does not specify what $\Lm$ should be. 

Neither Newton's constant $G$, nor Planck's constant $\hbar$ appear in the formulation of the quantum phase of the theory Eq. (\ref{eq:ERPI}). Therefore, none of them are fundamental constant in Entangled Relativity \cite{ludwig:2015pl,minazzoli:2022ar,minazzoli:2023ar}. From now on, we will refer to them as Newton's parameter or Planck's parameter. 
$\epsilon^2$ represents a quantum of energy squared, influencing the theory only at the quantum level. It serves the same role as the quantum of action, $\hbar$, in standard physics. To recover standard quantum field theory in the limit where gravity is negligible, one can show that $\epsilon$ corresponds to the reduced Planck energy \cite{minazzoli:2022ar,minazzoli:2023ar}, defined as $\epsilon \equiv \sqrt{\hbar c^5 / (8 \pi G)}$. Since neither $G$ nor $\hbar$ is constant in Entangled Relativity, this implies that $G \propto \hbar$ in the semi-classical limit where gravity is treated as a classical background \cite{minazzoli:2022ar,minazzoli:2023ar}, thereby establishing an explicit connection between the quantum and gravitational realms.

Classical physics corresponds to paths with stationary phases, $\delta \Theta = 0$, and is entirely insensitive to the value of $\epsilon$—just as classical physics in standard physics is insensitive to the value of the quantum of action, $\hbar$. Those paths correspond to the following field equations  \citep{ludwig:2015pl,minazzoli:2018pr}
\be \label{eq:metric}
G_{\mu \nu} = \kappa T_{\mu \nu} + f_R^{-1} \left[\nabla_\mu \nabla_\nu - g_{\mu \nu} \Box \right] f_R,
\ee
with
\bea
\kappa = - \frac{R}{\Lm},\qquad
f_R = \frac{1}{2 \epsilon^2} \frac{\Lm^2}{R^2} = \frac{1}{2 \epsilon^2 \kappa^2},\label{eq:f_Rkappa}
\eea
with the following stress-energy tensor
\be
T_{\mu \nu} := -\frac{2}{\sqrt{-g}} \frac{\delta\left(\sqrt{-g} \mathcal{L}_{m}\right)}{\delta g^{\mu \nu}},
\ee
which is not classically conserved
\be
\nabla_{\sigma}\left(\frac{\mathcal{L}_{m}}{R} T^{\alpha \sigma}\right)=\mathcal{L}_{m} \nabla^{\alpha}\left(\frac{\mathcal{L}_{m}}{R}\right). \label{eq:noconsfR}
\ee

In the particular case of eletromagnetism, using units such that the magnetic permeability of vacuum is $\mu_0=1$, one has
\be
\Lm = - \frac{1}{2} F_{\mu \nu} F^{\mu \nu},
\ee
with $F_{\mu \nu}=\partial_\mu A_\nu - \partial_\nu A_\mu$ and $A^\alpha$ the four-vector electromagnetic potential. The stress-energy tensor is
\be
T_{\mu \nu} = 2 \left(F_{\mu \sigma} F_\nu^{~\sigma} + g_{\mu \nu} \frac{\Lm}{2}\right),
\ee
and is such that $T=0$. The electromagnetic field equation on the other hand reads
\be
\nabla_\sigma \left(\frac{\Lm}{R} F^{\mu \sigma}\right) = 0.\label{eq:Maxwell}
\ee

\subsection{General relativistic limit}
\label{sec:gen_limit}

Whenever matter fields satisfy $\Lm = T$ on-shell, the solutions of General Relativity are also solutions of Entangled Relativity. This is due to an intrinsic decoupling originally identified in scalar-tensor theories \cite{minazzoli:2013pr}. Specifically, by taking the trace of the metric field equation (\ref{eq:metric}), one can verify that the theory possesses an additional gravitational scalar degree of freedom—similar to standard $f(R)$ theories \cite{capozziello:2015sc}—whose equation is given by \citep{ludwig:2015pl,minazzoli:2018pr} 
\be 
3f_R^{-1} \square f_R = \kappa \left(T - \Lm\right). \label{eq:fRmetricfieldt} 
\ee 
As a result, whenever $\Lm = T$ on-shell, $f_R = (2 \epsilon^2 \kappa_{GR}^2)^{-1}$ is a solution of the full field equations of Entangled Relativity---where $\kappa_{GR} \equiv 8 \pi G / c^4$ is the coupling constant of General Relativity---and one exactly recovers the equations of General Relativity minimally coupled to matter fields. In particular, when $\Lm = T$ on-shell, and according to Eq. (\ref{eq:f_Rkappa}), $f_R = (2 \epsilon^2 \kappa_{GR}^2)^{-1}$ corresponds simply to the trace of Einstein's metric equation—that is, $R = -\kappa_{GR} T$.

Moreover, as far as we know, our present universe can be well-approximated as being composed of dust and electromagnetic radiation, both of which imply $\Lm = T$. For dust, one has $\Lm = -c^2 \sum_i m_i \delta^{(3)}(\vec{x} - \vec{x}_i(t)) / (\sqrt{-g} u^0) = T$, where $m_i$ represents the conserved mass of dust particles along their geodesics \cite{minazzoli:2013pd}. Similarly, for electromagnetic radiation, $\Lm \propto E^2 - B^2 = 0 = T$. Consequently, one expects Entangled Relativity and General Relativity to exhibit similar phenomenologies in our present universe.
\section{Black-hole in a magnetic field}

We seek for a solution akin to the Schwarzschild-Melvin solution of General Relativity \cite{griffiths:2009bk}. The Schwarzschild-Melvin spacetime is an exact solution of the Einstein electrovacuum equations that describes a black hole immersed in a magnetic field asymptotically aligned with the z-axis. This solution is significant for understanding the interaction between geometry and matter and is frequently employed as a model for astrophysical environments \cite{cardoso:2024ar}. Using units such that 
\be
\lim_{A^\alpha \rightarrow 0} \kappa =1,
\ee
where $A^\alpha$ is the electromagnetic 4-vector,
the solution for the metric is
\bea \label{eq:solmag}
&&ds^2 =  \Lambda^{-\frac{28}{13}} r^2 sin^2(\theta) d\varphi^2\\
&&+\Lambda^{\frac{20}{13}} \left[\left(1 - \frac{\rs}{r}\right)^{-1}dr^2 + r^2 d\theta^2 - \left(1 - \frac{\rs}{r}\right) c^2dt^2 \right], \nonumber
\eea
where $\rs$ is the Schwarzschild radius and with
\be \label{eq:units}
\Lambda = 1 + \frac{13}{48} B^2 r^2 sin^2\theta.
\ee
The solution for the electromagnetic 4-vector is
\be
A =\frac{B}{2 \Lambda} r^2 \sin^2 \theta~ d\varphi.
\ee
It describes a magnetic field pointing along the z-direction \cite{griffiths:2009bk} that reads
\be
B^\mu = \frac{B r \sin \theta}{\Lambda^2}\left[0,- r  \cos \theta,  \sin \theta, 0\right].
\ee
One can check that one has $T=0$. Regarding the Petrov classification of the spacetime described by the metric (\ref{eq:solmag}), we find that the external part is algebraically general, just as the Schwarzschild-Melvin solution in General Relativity \cite{griffiths:2009bk}.
Let us note that the solution is such that
\be
\Lm =-  B^2 \Lambda^{-44/13} \left(1- \frac{\rs~\sin^2 \theta}{r}\right),
\ee
and
\be
R= B^2 \Lambda^{-46/13}\left(1- \frac{\rs~\sin^2 \theta}{r}\right),
\ee
\footnote{Whereas $R=0$ and $\Lm = - B^2 \Lambda^{-4} \left(1- \rs \sin^2 \theta/r\right)$ for the Schwarzschild-Melvin solution of General Relativity.} such that one has
\be \label{eq:ratiomag}
\frac{R}{\Lm} = -\Lambda^{-2/13}.
\ee
Therefore, it is clear from the onset that the magnetic solution is well defined in the $\Lm \rightarrow 0$ limit. It is also worthwhile to note that the solution is such that $\Lm<0$ and $R>0$, such that $R/\Lm<0$. In particular, for $\theta=0$, one has $\Lm = -B^2$ and $R=B^2$, such that $\kappa=1$.\footnote{The solution and its properties have been checked using SageManifolds \cite{gourgoulhon:2015jc}. The notebook is freely accessible at the following url : \url{https://github.com/ominazzoli/ER-Schwarzschild-Melvin/blob/main/ER\%20Melvin\%20magnetic\%20\%2B\%20Schwarzschild.ipynb}.}
\subsection{Limiting cases}

\subsubsection{$B \rightarrow 0$}

For $B \rightarrow 0$, one has $\Lambda \rightarrow 1$, such that the Schwarzschild black-hole is a limit of the solution Eq. (\ref{eq:solmag}) when the magnetic field disappears. In that limit, both $R$ and $\L_m$ tend to zero, but their ratio remains well defined in that limit, as one can see in Eq. (\ref{eq:ratiomag}). 
Given the existence of cosmological magnetic fields with extremely weak strengths in our universe, the solution Eq. (\ref{eq:solmag}) with $B \rightarrow 0$ is a good approximation of astrophysical (spherical) black holes. This is because astrophysical black holes have negligible charges due to their formation processes and their tendency to attract oppositely charged particles more than similarly charged ones.
At first, given that the field equations of Entangled Relativity are different from those of General Relativity, it may be surprising that one obtains a metric of General Relativity as a limiting case when the magnetic field disappears. However, this can be easily understood by taking the trace of the metric field equation.
Slightly reordering Eq. (\ref{eq:fRmetricfieldt}), it reads
\be
3\square \frac{\mathcal{L}_{m}^{2}}{R^{2}}=-\frac{\Lm}{R}\left(T-\Lm\right). \label{eq:fRmetricfieldt2}
\ee
This equation shows that the ratio between $\Lm$ and $R$ is a new scalar degree of freedom in Entangled Relativity \cite{ludwig:2015pl}. With a magnetic field, such a scalar degree of freedom is sourced because $\Lm \neq 0$ whereas $T=0$. However, when $(\Lm, T) \rightarrow 0$, the scalar degree of freedom is no longer sourced and remains constant. When this happens, one can check that the whole set of field equations becomes equivalent to those obtained in the framework of General Relativity minimally coupled to matter fields.

That being said, it would not have been possible to start finding a solution from the vacuum assumption—that is $(\Lm, T) = 0$—as the ratio between $R$ and $\Lm$ would not have been well-defined in Eq. (\ref{eq:metric}) in this case. Hence, we still argue that Entangled Relativity prevents the existence of vacuum solutions,\footnote{Moreover, as soon as $\Lm$ is defined in the path integral formulation of the theory using Eq. (\ref{eq:ERPI}) \cite{minazzoli:2022ar,minazzoli:2023ar}, there will always be a non-null effective background field composed of the quantum vacuum. The issue of the effective contribution of the quantum vacuum to the field equations in this theory is yet to be investigated.} but it does have the vacuum solutions of General Relativity as asymptotic solutions in the $(\Lm, T) \rightarrow 0$ limit.

\subsubsection{$\rs \rightarrow 0$}

When the size of the black-hole is set to zero, the metric becomes

\bea \label{eq:solmagm0}
ds^2 &=& \Lambda^{\frac{20}{13}} \left[dr^2 + r^2 d\theta^2 - dt^2 \right] \nonumber\\
&&+ \Lambda^{-\frac{28}{13}} r^2 sin^2(\theta) d\varphi^2. 
\eea

It deviates from the Melvin solution \cite{griffiths:2009bk} because the extra degree-of-freedom is sourced---see Eq. (\ref{eq:fRmetricfieldt})---such that the field equations deviate from the ones of General Relativity. Nevertheless, the structure is quite similar to the Melvin solution of General Relativity.
\subsubsection{$(B,\rs) \rightarrow 0$}

Quite interestingly, since $\lim_{B \to 0}\Lambda = 1$, we also find that Minkowski is a limit when both $B$ and $\rs$ tend to zero, for which the ratios $R/\Lm$ and $\Lm/R$ remain finite all along.


\section{Black-hole in an electric field}

The electric version of the solution can be obtained through the following transformation

\begin{subequations}\label{eq:transfo}
\bea
&&F_{\mu \nu}  \longrightarrow F^{e}_{\mu \nu} =-\frac{1}{2} \frac{\Lm}{R} ~ \epsilon_{\mu \nu \kappa \lambda} F^{\kappa \lambda},\\
&&g_{\mu \nu} \longrightarrow g^e_{\mu \nu} = \left(\frac{\Lm}{R}\right)^4 g_{\mu \nu}.
\eea
\end{subequations}\label{eq:litdiffm}
The resulting solution for the metric is
\bea\label{eq:solelec}
ds^2 &=& \Lambda^{-\frac{20}{13}} r^2 sin^2(\theta) d\varphi^2\\
&&+ \Lambda^{\frac{28}{13}} \left[\left(1 - \frac{\rs}{r}\right)^{-1}dr^2 \right.\nonumber \\
&& \left. + r^2 d\theta^2 - \left(1 - \frac{\rs}{r}\right) c^2dt^2 \right],\nonumber
\eea

with
\be
\Lambda = 1 + \frac{13}{48} E^2 r^2 sin^2\theta.
\ee
Whereas, the resulting solution for the electromagnetic 4-vector is

\be
A =E \left(1-\frac{\rs}{r} \right)r \cos \theta ~ dt.
\ee
It describes an electric field pointing along the z-direction \cite{griffiths:2009bk} that reads
\be
E^\mu = E \left[0, \cos \theta, - \left(1-\frac{r_S}{r}\right) r\sin \theta, 0\right].
\ee
One can check that one has $T=0$. We find that the external part ($r>\rs$) of the spacetime in Eq. (\ref{eq:solmag}) is algebraically general. This is expected, given that the magnetic and electric metric solutions differ by a conformal factor in Eq. (\ref{eq:transfo}), while the Petrov classification is invariant under conformal transformations. Let us note that the solution is such that
\be
\Lm = E^2 \Lambda^{-56/13} \left(1- \frac{\rs~\sin^2 \theta}{r}\right),
\ee
and
\be
R= - E^2 \Lambda^{-54/13} \left(1- \frac{\rs~\sin^2 \theta}{r}\right),
\ee
such that one has
\be \label{eq:ratioelec}
\frac{R}{\Lm} = -\Lambda^{2/13}.
\ee
Therefore, it is clear from the onset that the electric solution is also well defined in the $\Lm \rightarrow 0$ limit. It is also worthwhile to notice that the solution is such that $\Lm>0$ and $R<0$---that is, the opposite situation with respect to the magnetic case---such that one still has $R/\Lm<0$ nonetheless. In particular, for $\theta=0$, one has $\Lm = E^2$ and $R=-E^2$, such that $\kappa=1$.\footnote{The notebook that checks this solution and its properties can be found at the following address : \url{https://github.com/ominazzoli/ER-Schwarzschild-Melvin/blob/main/ER\%20Melvin\%20electric\%20\%2B\%20Schwarzschild.ipynb}.}
\subsection{Limiting cases}

As in the magnetic case, one obtains the Swcharzschild black-hole in the $E \rightarrow 0$ limit, and Minkowski in the $(E,\rs)\rightarrow 0$ limit. However, the solution without a black-hole is not the same as in Eq. (\ref{eq:solmagm0}), since it reads as follows
\bea \label{eq:solelecm0}
ds^2 &=& \Lambda^{\frac{28}{13}} \left[dr^2 + r^2 d\theta^2 - dt^2 \right] \nonumber\\
&&+ \Lambda^{-\frac{20}{13}} r^2 sin^2(\theta) d\varphi^2. 
\eea


\section{Discussion}


Before the present result, it was not known whether or not Entangled Relativity could have neutral black-hole solutions. Now we know this is the case for spherical black-holes, and we can expect that it should also be the case for rotating black-holes—although rotating neutral solutions remain to be found. As with the case of charged black-hole solutions studied in \cite{minazzoli:2021ej,wavasseur:2024ar}, we found that the vacuum black-hole solutions of General Relativity are limits of black-hole solutions of Entangled Relativity when the value of the matter field amplitude goes to zero. Since such a value is expected to be extremely small in the universe—whether one is talking about the charge of a black-hole, or the background density of magnetic or electric fields surrounding black-holes—the present results are yet another indication that vacuum black-hole solutions of General Relativity are good approximations of black-holes in Entangled Relativity. Therefore, it presently seems unlikely to us that one may be able to discriminate between the two theories through the observations of black-holes and associated phenomena. Neutron stars seem to be better targets in that respect \cite{arruga:2021pr,arruga:2021ep}.

One interesting point with the solutions presented in this paper is that they are such that the matter Lagrangian is either positive ($\Lm \propto E^2$) or negative ($\Lm \propto -B^2$) depending on whether one is dealing with an electric or magnetic field. This is interesting because one can see in Eq. (\ref{eq:f_Rkappa}) that the sign of the gravitational coupling parameter $\kappa$ is proportional to the ratio between $\Lm$ and $R$. Indeed, Entangled Relativity does not prevent, \textit{a priori}, a negative value the gravitational parameter $\kappa$. This would, however, be quite inconsistent with any observations so far, as a negative parameter would imply repulsive gravity. Furthermore, one might na\"ively expect the sign of $\kappa$ to change with a change in the sign of $\Lm$, and the definition of $\kappa$ to be singular when $\Lm \rightarrow 0$.\footnote{These arguments have notably been used by a referee to prevent the publication of \cite{arruga:2021ep} in \textit{Phys. Rev. D}.} But for the solutions found in the present paper—as well as for the charged black-hole solutions \cite{wavasseur:2024ar}—this is not the case. This is due to the fact that when $\Lm$ changes sign or goes to zero, so does $R$, such that the ratio is always well-defined and the sign of $\kappa$ remains the same in all cases.
Given our present understanding of Entangled Relativity, we do not expect $\kappa$ to vary much in the observable universe. Indeed, it has been found that it only varies by a few percent within the densest objects of the observable universe \cite{arruga:2021pr,arruga:2021ep}. However, it cannot be excluded \textit{a priori} that the value of $\kappa$ varies much more widely in the collapsing dense compact object inside a black-hole, that is, in the collapsing object after the formation of the event horizon. Furthermore, we currently do not see any reason to exclude \textit{a priori} that $\kappa$ might even have its sign changed, potentially leading to repulsive gravity that could prevent the formation of a singularity. One could use the same line of argument for the primordial universe. These prospects are fascinating future avenues of research in Entangled Relativity.

\section*{Data Availability Statement}

No Data associated in the manuscript.

\bibliography{ER_SMelvin}

\end{document}